\documentclass[10pt,conference]{IEEEtran}
\IEEEoverridecommandlockouts

\usepackage[noadjust]{cite}
\usepackage{amsmath,amssymb,amsfonts}
\usepackage{algorithmic}
\usepackage{graphicx}
\usepackage{textcomp}
\usepackage{xcolor}
\usepackage[utf8]{inputenc}
\usepackage{newunicodechar,graphicx}
\usepackage[hidelinks,bookmarks=true]{hyperref}
\usepackage[numbered]{bookmark}
\usepackage{soul}
\usepackage{booktabs}
\usepackage{multirow}
\usepackage{float}
\usepackage{url}
\usepackage{tcolorbox}
\usepackage{tabularx} 
\usepackage[switch]{lineno}
\usepackage{enumitem}
\usepackage{seqsplit}
\usepackage{pifont}
\usepackage{balance}
\usepackage{microtype}

\hypersetup{
	pdfborder          = {0 0 0},
	breaklinks         = true,
	bookmarksnumbered  = true,
	pdfsubject         = {6th International Workshop on Refactoring (IWoR 2022)},
	pdfkeywords        = {Software Quality, Refactoring, Rename Refactoring, Program Comprehension, Identifier names, Empirical Study},
	pdftitle           = {Rename Chains:  An Exploratory Study on the Occurrence and Characteristics of Identifiers Undergoing Multiple Renamings},
	pdfauthor          = {Anthony Peruma, Christian D. Newman}
}
\usepackage{listings}

\definecolor{javared}{rgb}{0.6,0,0} 
\definecolor{javagreen}{rgb}{0.25,0.5,0.35} 
\definecolor{javapurple}{rgb}{0.5,0,0.35} 
\definecolor{javadocblue}{rgb}{0.25,0.35,0.75} 
 
\definecolor{light-red}{rgb}{1,0.92,0.91}
\definecolor{light-green}{rgb}{0.9,1,0.93}
\DeclareRobustCommand{\okina}{%
  \raisebox{\dimexpr\fontcharht\font`A-\height}{%
    \scalebox{0.8}{`}%
  }%
}
\newunicodechar{ʻ}{\okina}
\newcommand{\RQA}{\textbf{RQ1}: To what extent do identifiers undergo multiple rename refactoring operations?}
\newcommand{\RQB}{\textbf{RQ2}: How frequently do renames occur within a rename chain, and who is responsible for their creation?}
\newcommand{\RQC}{\textbf{RQ3}: How do the semantics of an identifier's name evolve in a rename chain?}
\newcommand{\RQD}{\textbf{RQ4}: To what extent can commit log messages help contextualize the occurrence of rename chains?}

\begin{document}

\title{Rename Chains:  An Exploratory Study on the Occurrence and Characteristics of Identifiers Undergoing Multiple Renamings}


\author{\IEEEauthorblockN{Anthony Peruma}
\IEEEauthorblockA{\textit{Information and Computer Sciences Department} \\
\textit{University of Hawaiʻi at Mānoa}\\
Honolulu, Hawaiʻi, USA \\
peruma@hawaii.edu}
\and
\IEEEauthorblockN{Christian D. Newman}
\IEEEauthorblockA{\textit{Department of Software Engineering} \\
\textit{Rochester Institute of Technology}\\
Rochester, New York, USA \\
cnewman@se.rit.edu}
}

\maketitle

\begin{abstract}
Identifier names play a significant role in program comprehension activities, with high-quality names improving developer productivity and system quality. To correct poor-quality names, developers rename identifiers to reflect their intended purpose better. However, renames do not always result in high-quality, long-lasting names; in many cases, developers perform multiple rename operations on the same identifier throughout the system's lifetime. In this paper, we report on a large-scale empirical study that examines the occurrence of identifiers undergoing multiple renames (i.e., rename chains). Our findings show the presence of rename chains in almost every project, with methods typically having more rename chains than other identifier types. Furthermore, it is usually the same developer responsible for creating all renames within a chain, with most names maintaining the same grammatical structure. Understanding rename chains can help us provide stronger advice, and targeted research, on how to craft high-quality, long-lasting identifiers.

\end{abstract}


\section{Introduction}
Be it bug fixing or updating features, program comprehension is an essential part of any software maintenance activity \cite{Rajlich2002WPC}. Program comprehension is the act of developers reading the code to understand its behavior in order to know where to make updates to the source code \cite{Mayrhauser1995Computer}. Therefore, to ensure both developer productivity and system quality, it is essential for developers to craft identifiers with meaningful names. In other words, the name should accurately reflect its intended behavior.

Research shows that identifier names account for almost 70\% of the characters in a software system’s codebase \cite{Deissenboeck2006SQJ}, with well-constructed names improving comprehension activities by an estimated 19\% \cite{Hofmeister2017SANER}. Unfortunately, there are significant problems with many identifiers, and no generalizable methods to measure identifier quality. This is likely part of the reason renaming is one of the most frequent types of rework (i.e., refactoring) developers perform on their code base, contributing to around 40\% of the rework developers perform throughout the lifetime of the system \cite{Peruma2020JSS, Peruma2018IWoR, Peruma2020IWoR}. 

While Rename Refactoring is the approach developers take to correct poor-quality names, there is no guarantee that the resulting new name is of high quality, with developers sometimes performing multiple rename operations to the same identifier. For instance, let us compare the code snippets in Listing \ref{Listing:introExample01} and Listing \ref{Listing:introExample02}, both of which show multiple renamings of a method's name. In Listing \ref{Listing:introExample01}, the developer renames the method \texttt{\seqsplit{sendPacket2$\rightarrow$sendPacket3$\rightarrow$syncedSendPacket}}. The original and first iteration of the name contain digits, and this type of naming is known as a \textit{Distinguisher} \cite{Peruma2022NLBSE}. Developers utilize such names to prevent name collisions at compile time when multiple identifiers with the same name are in the class/file. The final version of the name, syncedSendPacket, is no longer a \textit{Distinguisher} and is more descriptive than the original. In contrast, it can be argued that the end result of the method rename in Listing \ref{Listing:introExample02} does not produce a high-quality name as \texttt{println} is a copy of the statement inside the method and provides no additional information.

Even though prior work on identifier naming examines the lexical semantic updates developers make to a name when performing the rename operation \cite{Arnaoudova2014TSE,Peruma2018IWoR,Peruma2020JSS,Peruma2021ICPC}, they fall short of investigating how each identifier evolves throughout the system's entire lifetime. In other words, they do not examine if the individual rename operations are related to each other. Likewise, studies that propose rename opportunities in the code do not consider the historical evolution of the identifiers \cite{Allamanis2014FSE,Allamanis2015FSE,Suzuki2014ICPC}.

\vspace{3mm}
\begin{lstlisting}[caption=An example of a method name undergoing multiple renames to make it more descriptive of its purpose (\cite{introExample01_a}$\rightarrow$\cite{introExample01_b})., label=Listing:introExample01, firstnumber = last, escapeinside={(*@}{@*)},escapechar=!]
!\colorbox{light-red}{- public void sendPacket2(Packet9Respawn packet) \{}!
!\colorbox{light-green}{+ public void sendPacket3(Packet9Respawn packet) \{}!
!\colorbox{light-red}{- public void sendPacket3(Packet9Respawn packet) \{}!
!\colorbox{light-green}{+ public void syncedSendPacket(Packet9Respawn packet) \{}!
    activeChunks.clear();
	super.sendPacket(packet);
}
\end{lstlisting}

\newpage
\begin{lstlisting}[caption=An example of a rename chain resulting in a weak method name; it is just a copy of the statement within the method (\cite{introExample02_a}$\rightarrow$\cite{introExample02_b})., label=Listing:introExample02, firstnumber = last, escapeinside={(*@}{@*)},escapechar=!]
!\colorbox{light-red}{- public void writeMessage(String message) \{}!
!\colorbox{light-green}{+ public void info(String message) \{}!
!\colorbox{light-red}{- public void info(String message) \{}!
!\colorbox{light-green}{+ public void println(String message) \{}!
     systemOut.println(message);
}
\end{lstlisting}

\subsection{Goal \& Research Questions}
The goal of this study is to explore the evolution of identifier names by \textit{constructing and studying the characteristics of a chain of renames for identifiers (i.e., a rename chain)}. Through the findings from our study, we aim to understand the multiple rename refactoring operations developers perform on an identifier that can feed into tools and techniques to better support developers with crafting and maintaining identifiers in their code. Therefore, we propose and answer the following research questions (RQs): 

\vspace{1.3mm}
\noindent\textbf{\RQA} 
This RQ reports on the volume and types of identifiers that undergo multiple renames during their lifetime and how frequently they occur in projects. Knowing the popularity of rename chains in a project's evolution will direct us to further research in this area.

\vspace{1mm}
\noindent\textbf{\RQB} 
From this RQ, we gain insight into the developers performing the renames in the chain and how frequently developers perform the rename operations within the chains. By considering the developers responsible for creating rename chains, rename recommendation techniques can improve their accuracy and usability.

\vspace{1mm}
\noindent\textbf{\RQC}
This RQ analyzes the lexical-semantic properties of the renames by comparing the part-of-speech tags of the first and last names in the chain. Findings and heuristics from this RQ can be incorporated into automated identifier name appraisal and recommendation tools and techniques.

\vspace{1mm}
\noindent\textbf{\RQD}
Using the generated rename chains in our dataset, this RQ examines how effectively commit messages can identify the specific causes for developers to create rename chains.

\subsection{Contribution}
The main contributions from this work are as follows:
\begin{itemize}
    \item Our results represent a significant step toward understanding how an identifier's name evolves through the project's lifetime. Through our discussion, we pave the way for subsequent research to enhance our knowledge of high-quality identifier naming, especially in automated identifier name appraisal and recommendation tools.
    \item We make our dataset of rename chains, including specific characteristics of the renames, publicly available.
\end{itemize}

\section{Related Work}
\label{Section:related_work}
This section discusses the work related to identifier renaming. Broadly, these studies fall into two categories, empirical studies that examine the semantic characteristics of names and studies that propose rename recommendation techniques/models.

\subsection{Empirical Studies}
In a developer survey, Arnoudova et al. \cite{Arnaoudova2014TSE} report that developers perform renaming as part of their implementation workflow and admit that renaming is not straightforward. Furthermore, the authors propose a taxonomy to classify the types of semantic updates a name undergoes when renamed. 

An empirical examination by Peruma et al. \cite{Peruma2018IWoR} of the semantic updates developers make to a name shows that developers frequently make simple renames by adding or removing a single term in a name. Further, the authors also show that developers frequently narrow the meaning of the name. The authors also highlight specific grammar patterns developers utilize when crafting unit test method names \cite{Peruma2021ICPC} and produce a taxonomy of digits occurring in an identifier's name \cite{Peruma2022NLBSE}. Additionally, as a means of contextualizing the renames developer perform, Peruma et al. \cite{Peruma2020JSS} show relationships between the data type and the plurality of the name. Specifically, the name changes from singular to plural when the data type changes from a non-collection to a collection type. The authors also show that specific identifier renamings tend to co-occur with other types of refactoring operations. Further, the authors also show that novice developers tend to perform more renames than other types of refactoring operations \cite{Peruma2019SCAM}.

\subsection{Rename Recommendations}
A model called NATURALIZE that uses statistical natural language processing to mine and learn the style (i.e., coding norms) of a codebase and offers renaming recommendations are introduced by Allamanis et al. in their paper \cite{Allamanis2014FSE}. To standardize names used in related contexts, NATURALIZE learns syntactic restrictions, or sub-grammars, on identifier names like camelcase or underscore. The authors also recommend a neural probabilistic language model to automatically suggest descriptive, idiomatic method and class names \cite{ Allamanis2015FSE}. An n-gram based approach for assessing the comprehensibility of method names and recommending intelligible method names is introduced by Suzuki et al. \cite{Suzuki2014ICPC}. The authors' solution involves gathering and learning method names from Java systems. The authors employ the n-gram model to provide recommendations to the developer and a threshold to assess the comprehensibility score of a method's name as part of their analysis process. Deep learning methods are used by Liu et al. \cite{Liu2019ICSE} to spot incorrect method names. Their methodology retrieves in-depth representations of method bodies and names. The model is trained by the authors using numerous techniques from actual projects. The name recommendation method compares the overlap between the set of method names whose bodies are close in the method body vector space and the closeness of method names in the method name vector space.

\section{Experiment Design}
\label{Section:experiment_design}
\begin{figure*}[t]
 	\centering
 	\includegraphics[trim=0cm 0cm 0cm 0cm,clip,scale=0.66]{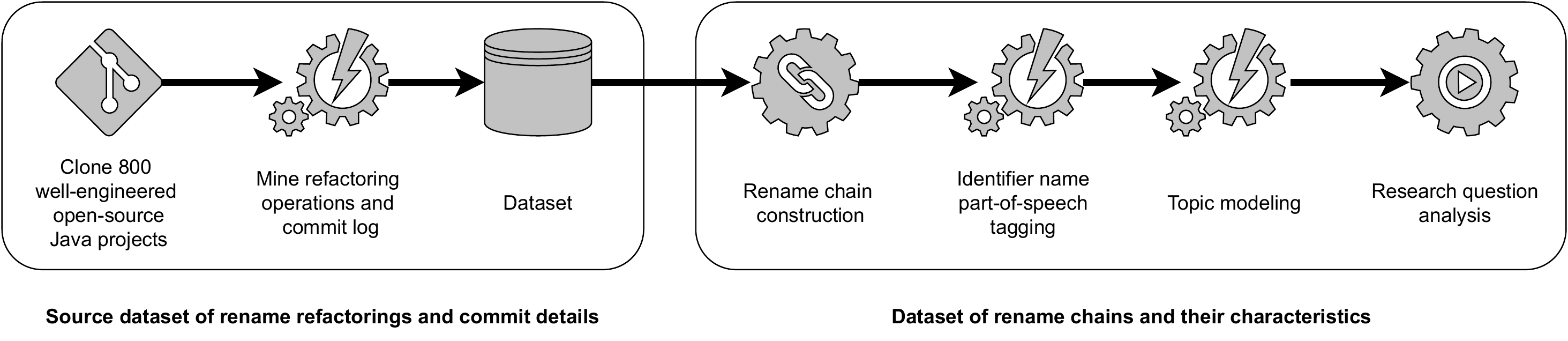}
 	\caption{Overview of our experiment design.}
 	\label{Figure:diagram_experiment}
\end{figure*}

In this section, we provide details about the methodology for our study. Figure \ref{Figure:diagram_experiment} shows a high-level overview of our experiment, which we describe in detail below. Furthermore, the dataset we utilize/generate in this study is available on our project website for replication and extension purposes \cite{website}. 

\subsection{Source Dataset}
In this study, we utilize an existing dataset of mined refactoring operations made available by Peruma et al. \cite{Peruma2021ICPC} from their research on test method name renaming and resued by the authors in another study on identifier names \cite{Peruma2022NLBSE}. The authors of the dataset utilized RefactoringMiner \cite{Tsantalis2018ICSE} to mine the refactoring operations of 800 well-engineered open-source Java systems. RefactoringMiner is a state-of-the-art tool that iterates through a project's commit log mining refactoring operations. The dataset contains mined rename refactoring operations for classes, attributes, methods, parameters, and variables. 

\subsection{Rename Chain Construction}
Our manual analysis of the source dataset shows the presence of auto-generated code related to projects utilizing Antlr. Since such code could skew our findings, we first exclude such source files from our analysis; our dataset contains the query we used to perform the exclusion. After performing this exclusion, we start the work of constructing the rename chain for each identifier type, using custom scripts. Our approach involves the use of the fully qualified name of the identifier to perform name comparisons to form links in the chain. The general approach involves first obtaining the refactorings for each identifier type sorted by the author commit date for each project. Next, for each type of identifier rename refactoring in the project, we search for instances where the new name in the refactoring is the old name in a subsequent rename operation. If such a match exists, it forms a link in the chain. This process continues recursively. Our replication package contains the code utilized to create the chains for each identifier type. 

\subsection{Part-of-Speech Tagging}
To understand the semantic change to an identifier's name, we utilize a specialized identifier name part-of-speech tagger made available by Newman et al. \cite{Newman2021TSE}. This is a state-of-the-art tagger that outperforms other taggers, including the Stanford tagger \cite{Toutanova2003feature}, for identifier names. The tagger utilizes a subset of the Penn Treebank tagset \cite{Marcinkiewicz1994building} and includes nouns, verbs, noun modifiers, determiners, etc; details of which are available at \cite{SCANL-Catalog}. Using this tagger, we generate the part-of-speech tags for each term in an identifier's name for the original and last name in the rename chain.

\subsection{Topic Modeling}
To contextualize the presence of rename chains, we perform a topic modeling analysis utilizing the latent Dirichlet allocation (LDA) algorithm \cite{blei2003latent}. Before performing the topic modeling analysis, we perform a set of text preprocessing tasks on the commit messages; we remove non-alphabet characters, such as punctuations, set the text to lowercase, lemmatize words, and finally, remove standard English stopwords. To arrive at the optimal number of topics, we iteratively extracted topics from two to ten in increments of one, where each topic execution cycle had 100 passes and 200 iterations. We manually examined the word frequencies present in each topic cycle to determine the optimum topic. 

\subsection{Research Question Analysis}
To answer our research questions, we follow a mixed methods approach, where we supplement our quantitative findings with qualitative examples from our dataset. This technique helps us understand our results through contextualization. We employ custom scripts and database queries to answer our research questions and elaborate on our approach when addressing each research question in Section \ref{Section:experiment_results}.

\begin{table}
\centering
\caption{Volume of mined rename operations for each identifier type.}
\label{Table:RenameOperations}
\begin{tabular}{@{}lrr@{}}
\toprule
\multicolumn{1}{c}{\textbf{\begin{tabular}[c]{@{}c@{}}Identifier\\ Type\end{tabular}}} & \multicolumn{1}{c}{\textbf{\begin{tabular}[c]{@{}c@{}}Total Rename \\ Refactoring Operations\end{tabular}}} & \multicolumn{1}{c}{\textbf{Percentage}} \\ \midrule
Method                                                                              & 84,321                                                                                                     & 29.50\%                                 \\
Parameter                                                                               & 72,972                                                                                                      & 25.53\%                                 \\
Variable                                                                                 & 62,059                                                                                                       & 21.75\%                                  \\
Attribute                                                                                  & 35,741                                                                                                       & 12.51\%                                  \\
Class                                                                              & 30,693                                                                                                       & 10.74\%                                  \\
\textit{All}                                                                           & \textit{285,786}                                                                                            & \textit{100\%}                          \\ \bottomrule
\end{tabular}
\end{table}

\section{Experiment Results}
\label{Section:experiment_results}
In this section, we report on the findings of our experiments. Since our work is based on the rename refactoring operations performed on identifiers, we first present an overview of renames in our dataset. In total, our dataset contains 285,786 rename refactorings spread across the five identifier types. Table \ref{Table:RenameOperations} shows the volume of renames by identifier type, most of which were method renames (29.50\%).

Moving on, we focus our analysis on rename chains by answering our RQs. The first RQ examines the volume of rename chains present in the dataset, while The second and third RQ investigates specific characteristics of these rename chains. Due to space constraints, specific tables in the RQs show only the most frequently occurring instances; the complete set is available on our project website \cite{website}.

\subsection{\RQA}
In this RQ, we quantitatively analyze the rename chains in our dataset. In total, we mined 285,786 rename refactoring operations. We then analyzed this raw data to construct rename chains. A chain is a combination of rename operations applied to a single identifier. We construct a chain if two or more rename operations are applied to an identifier. In total, we detected 17,404 rename chains spread across all identifier types. In contrast, our dataset contains 247,567 identifiers that underwent only a single rename operation and, hence do not form a chain. A granular examination shows that, out of all identifier types in our dataset, methods (approx. 30.73\%) are most likely to have rename chains, followed by variables (approx. 23.47\%), classes (16.85\%), parameters (approx. 16.81\%), and attributes (approx. 12.14\%). Table \ref{Table:ChainCount} provides an overview of the mined rename chains.  

\begin{table}
\centering
\caption{Volume of identifiers with a single rename and multiple rename instances for each identifier type.}
\label{Table:ChainCount}
\begin{tabular}{@{}lrrr@{}}
\toprule
\multicolumn{1}{c}{\multirow{2}{*}{\textbf{\begin{tabular}[c]{@{}c@{}}Identifier\\ Type\end{tabular}}}} & \multicolumn{1}{c}{\multirow{2}{*}{\textbf{\begin{tabular}[c]{@{}c@{}}Instances with a\\ single rename\end{tabular}}}} & \multicolumn{2}{c}{\textbf{\begin{tabular}[c]{@{}c@{}}Instances with more than one rename\\ (i.e., rename chain)\end{tabular}}} \\
\multicolumn{1}{c}{}                                                                                    & \multicolumn{1}{c}{}                                                                                                   & \multicolumn{1}{c}{\textbf{Count}}                           & \multicolumn{1}{c}{\textbf{Percentage}}                          \\ \midrule
Class                                                                                                   & 24,221                                                                                                                  & 2,933                                                          & 16.85\%                                                           \\
Attribute                                                                                               & 31,245                                                                                                                 & 2,113                                                      & 12.14\%                                                          \\
Method                                                                                                  & 72,884                                                                                                                  & 5,349                                                          & 30.73\%                                                           \\
Parameter                                                                                               & 66,015                                                                                                                  & 2,925                                                           & 16.81\%                                                           \\
Variable                                                                                                & 53,202                                                                                                                 & 4,084                                                        & 23.47\%                                                          \\
\textit{All}                                                                                            & \textit{247,567}                                                                                                        & \textit{17,404}                                              & \textit{100.00\%}                                                \\ \bottomrule
\end{tabular}
\end{table}

Moving on, we focus on the number of rename operations that form a rename chain. Overall, an identifier rename chain contains a median of 2 and an average of 9 rename instances. On a granular level, we observe that classes, attributes, methods, and parameters have a median of two renames in their chains, while variables have three rename instances. Table \ref{Table:StatSummary_Chain} shows a statistical summary of the number of rename instances for each identifier type in their rename chain.

\begin{table}
\centering
\caption{Statistical summary of the number of rename instances associated with each type of identifier.}
\label{Table:StatSummary_Chain}
\begin{tabular}{@{}crrrrr@{}}
\toprule
\textbf{Min.}         & \multicolumn{1}{c}{\textbf{1st Qu.}} & \multicolumn{1}{c}{\textbf{Median}} & \multicolumn{1}{c}{\textbf{Mean}} & \multicolumn{1}{c}{\textbf{3rd Qu.}} & \multicolumn{1}{c}{\textbf{Max.}} \\ \midrule
\multicolumn{6}{c}{\textit{Class}}                                                                                                                                                                                \\
\multicolumn{1}{r}{2} & 2                                    & 2                                   & 2.23                              & 2                                    & 7                                \\ \midrule
\multicolumn{6}{c}{\textit{Attribute}}                                                                                                                                                                            \\
\multicolumn{1}{r}{2} & 2                                    & 2                                   & 2.16                              & 2                                    & 6                                \\ \midrule
\multicolumn{6}{c}{\textit{Method}}                                                                                                                                                                               \\
\multicolumn{1}{r}{2} & 2                                    & 2                                   & 2.17                              & 2                                    & 8                                \\ \midrule
\multicolumn{6}{c}{\textit{Parameter}}                                                                                                                                                                     \\
\multicolumn{1}{r}{2} & 2                                    & 2                                   & 2.49                              & 2                                    & 53                                 \\ \midrule
\multicolumn{6}{c}{\textit{Variable}}                                                                                                                                                                      \\
\multicolumn{1}{r}{2} & 2                                    & 3                                   & 31.09                              & 55                                    & 143                                \\ \bottomrule
\end{tabular}
\end{table}

Finally, while our dataset contains 798 projects having rename refactorings, 668 (or 83.71\%) of these projects contain rename chains. Looking at the volume of rename chains within these projects, we observe that projects have a median of nine and a mean of 26.05 identifiers undergoing multiple renames.

\begin{tcolorbox}[top=0.5pt,bottom=0.5pt,left=1pt,right=1pt]
\textbf{Summary for RQ1.}
Though rename operations are prevalent in the implementation and maintenance of software systems, most identifiers typically undergo a single rename throughout their lifetime. However, rename chains are present in most systems. Method names typically undergo multiple renamings and typically contain around two renames in their rename chain. Variables, on the other hand, undergo around three renamings.
\end{tcolorbox}

\subsection{\RQB}
In the prior RQ, we show the occurrence of rename chains in the evolution of the code base of a software system. Moving on, this RQ examines the renames occurring within these chains. More specifically, we investigate the interval duration between the renames in the chains and the developers performing these renames. The findings from this RQ help us better understand the characteristics of rename chains.

\noindent\textbf{Interval Analysis}

This analysis examines the interval (i.e., time duration) between renames in chains having two or more rename instances. An overall examination of the median number of days between renames shows that the renames occur two days apart. Next, in a more granular examination, we observe that attributes have a median of 25 days, followed by classes having 19 days, methods with 14 days, parameters with seven days, and variables having two days between renames in the chain. 

Our subsequent examination looks at the interval between the first and last rename in the chain. Parameters have the lowest interval with a median of 17 days between the first and last rename. In contrast, variables have the longest of 357 days between the first and last rename in the chain. Finally, classes, attributes, and methods have an interval of 32, 35, and 22 days, respectively.

\noindent\textbf{Developer Analysis}

In this analysis, we investigate who performs the renames involved in the rename chain. To this extent, we utilize the email address associated with the commit containing the rename (i.e., git author email). Prior studies have used the email address to determine unique developers, including those that examine identifier renaming \cite{Peruma2020JSS}. Our analysis is on chains having two or more rename instances.

First, the same developer performs just over half of the rename chains (i.e., 10,799 or 62.05\% instances). Next, focusing on the chains having multiple developers, we observe that 760 or 11.51\% of instances have a different developer performing the first and last rename in the chain. Furthermore, these multi-developer chains have a median and average of approximately two unique developers performing the renames in the chain. At a more granular level, attribute chains have the most developers involved in the rename process, with a median of four developers, followed by variables with a median of three. Class and method chains have a median of two developers performing the renames in their respective chains. 

\begin{tcolorbox}[top=0.5pt,bottom=0.5pt,left=1pt,right=1pt]
\textbf{Summary for RQ2.}
Rename chains are typically constructed with rename refactoring operations that occur days apart, with variables typically having the shortest duration (approx. two days) and attributes the longest. Furthermore, rename chains are usually constructed by the same developer. Finally, multi-developer chains usually involve two developers, with the construction of attribute chains involving more developers than other identifier rename chains. 
\end{tcolorbox}

\subsection{\RQC}
This RQ continues our analysis of the evolution of renames chains by examining the lexical-semantic structure of the identifier names in the chain. Our analysis includes examining the part-of-speech tags instead of the semantics of actual word since the tags are more constrained and leave less room for misinterpretation. Furthermore, prior work has shown that developers utilize specific grammatical patterns when crafting identifier names \cite{Newman2020JSS}. To this extent, we utilize a specialized ensemble tagger for identifier names (\cite{Newman2021TSE}) to generate the part-of-speech tags for the words in an identifiers name. 

Since a chain can be composed of a varying number of renames, comparing and analyzing each and every rename within the chain is not feasible. Hence, we limit our analysis to the first and last rename in the chain. In other words, we compare the name of the identifier before the first rename and the name of the identifier after the final rename. Our analysis shows that 7,266 or 41.75\% rename chains have the same part-of-speech pattern for the original and final name. 

Next, we examine the common part-of-speech patterns utilized for the original and final names for each identifier type. Shown in Table \ref{Table:PoSfirstlast} are the top three widely used part-of-speech tags for the original and final name for each identifier type. From this table, we observe that the majority of the commonly used part-of-speech tags for both names are the same. For example, in the class rename chain \texttt{\seqsplit{TestServlet$\rightarrow$TheTestServlet$\rightarrow$TestServlet}} (\cite{pos_pattern_a}$\rightarrow$\cite{pos_pattern_b}), the part-of-speech pattern starts with NM|N, then changes to DT|NM|N when the developer prepends the determiner ``The'' to the name, before finally reverting the name structure to NM|N. The complete set of part-of-speech patterns are available in our shared dataset. 

Furthermore, it is encouraging to note that developers utilize standard naming structures when crafting names for identifiers \cite{Newman2020JSS}. From Table \ref{Table:PoSfirstlast}, we can see that classes, attributes, parameters, and variables begin with either a noun/noun-plural (N/NPL) or noun modifier (NM), while methods start with a verb (V). Additionally, we also observe instances where developers correct poorly structured names. For example, in the attribute rename chain \texttt{\seqsplit{setToValue$\rightarrow$groupName$\rightarrow$groupNameTextArea}} (\cite{pos_pattern2_a}$\rightarrow$\cite{pos_pattern2_b}), the original name starts with a verb (i.e., ``set''), which is generally incorrect for an attribute. However, within the chain, the developer changes the name to start with a noun modifier

\begin{table}
\centering
\caption{Common part-of-speech patterns associated with the original and final name in the rename chain for each identifier type.}
\label{Table:PoSfirstlast}
\begin{tabular}{@{}llrr@{}}
\toprule
\multicolumn{2}{c}{\textbf{\begin{tabular}[c]{@{}c@{}}Part-of-Speech\\ Pattern\end{tabular}}} & \multicolumn{1}{c}{\multirow{2}{*}{\textbf{Count}}} & \multicolumn{1}{c}{\multirow{2}{*}{\textbf{Percentage}}} \\
\multicolumn{1}{c}{\textbf{Original Name}}      & \multicolumn{1}{c}{\textbf{Final Name}}     & \multicolumn{1}{c}{}                                & \multicolumn{1}{c}{}                                     \\ \midrule
\multicolumn{4}{c}{\textit{Class}}                                                                                                                                                                             \\
NM|NM|N                                         & NM|NM|N                                     & 309                                                 & 11.74\%                                                  \\
NM|N                                            & NM|N                                        & 216                                                 & 8.20\%                                                   \\
NM|NM|NM|N                                      & NM|NM|NM|N                                  & 172                                                 & 6.53\%                                                   \\
\multicolumn{2}{l}{\textit{Other Patterns}}                                                   & 1,936                                               & 73.53\%                                                  \\ \midrule
\multicolumn{4}{c}{\textit{Attribute}}                                                                                                                                                                         \\
NM|N                                            & NM|N                                        & 285                                                 & 13.87\%                                                  \\
N                                               & N                                           & 231                                                 & 11.24\%                                                  \\
NM|NM|N                                         & NM|NM|N                                     & 105                                                 & 5.11\%                                                   \\
\multicolumn{2}{l}{\textit{Other Patterns}}                                                   & 1,434                                               & 69.78\%                                                  \\ \midrule
\multicolumn{4}{c}{\textit{Method}}                                                                                                                                                                            \\
V|NM|N                                          & V|NM|N                                      & 401                                                 & 9.65\%                                                   \\
V|N                                             & V|N                                         & 273                                                 & 6.57\%                                                   \\
V                                               & V                                           & 217                                                 & 5.22\%                                                   \\
\multicolumn{2}{l}{\textit{Other Patterns}}                                                   & 3,266                                               & 78.57\%                                                  \\ \midrule
\multicolumn{4}{c}{\textit{Parameter}}                                                                                                                                                                         \\
N                                               & N                                           & 804                                                 & 27.94\%                                                  \\
NM|N                                            & NM|N                                        & 500                                                 & 17.37\%                                                  \\
N                                               & NM|N                                        & 216                                                 & 7.51\%                                                   \\
\multicolumn{2}{l}{\textit{Other Patterns}}                                                   & 1,358                                               & 47.19\%                                                  \\ \midrule
\multicolumn{4}{c}{\textit{Variable}}                                                                                                                                                                          \\
N                                               & N                                           & 1,202                                               & 32.93\%                                                  \\
NPL                                             & N                                           & 617                                                 & 16.90\%                                                  \\
NM|N                                            & NM|N                                        & 252                                                 & 6.90\%                                                   \\
\multicolumn{2}{l}{\textit{Other Patterns}}                                                   & 1,579                                               & 43.26\%                                                  \\ \bottomrule
\end{tabular}
\end{table}

A high-level examination of the words making up the name in the identifiers shows that there are 7,584 instances where the original and final names contain an equal number of words. Further, our dataset contains 3,901 chains with identical original and final names. Additionally, we encounter 38 rename chains where the only difference between the names is a change in the case (e.g., \texttt{\seqsplit{experimentEngine$\rightarrow$junkEngine$\rightarrow$ExperimentEngine}} \cite{pos_pattern3_a}$\rightarrow$\cite{pos_pattern3_b}) and seven chains where the difference is a removal/addition of punctuation(s) (e.g., \texttt{\seqsplit{\_locator$\rightarrow$loader$\rightarrow$locator}} \cite{pos_pattern4_a}$\rightarrow$\cite{pos_pattern4_b}). 

\begin{tcolorbox}[top=0.5pt,bottom=0.5pt,left=1pt,right=1pt]
\textbf{Summary for RQ3.}
There are numerous instances where even though the words in an identifier's name change, the grammatical structure of the initial and last name in the chain remains the same. Furthermore, developers frequently follow well-established identifier naming structures when crafting names.   
\end{tcolorbox}

\subsection{\RQD}
While the prior RQs examine the occurrence and characteristics of rename chains, we need to understand why developers create these chains. Since surveying all the developers responsible for creating chains in our dataset is not feasible, this RQ performs an automated analysis of the commit log messages associated with commits that form rename chains. We analyze the commit message from the second rename onwards for each rename chain (i.e., two or more renames) as the second rename indicates the start of the rename chain. In our analysis, we perform a topic modeling analysis utilizing the LDA algorithm, as described in Section \ref{Section:experiment_design}. The results of our LDA analysis yield three distinct topics associated with these messages -- Code Cleanup, Refactoring, and Bug Fix/Testing.

The Code Cleanup topic includes words such as `renaming', `naming', `convention', `cleanup', and `whitespace', where the renames in the chain are due to the developer improving code style quality by adhering to standards, which includes following naming standards.  For example, in the chain \texttt{\seqsplit{myFilenameFilter$\rightarrow$libFilenameFilter$\rightarrow$LibFilenameFilter}} (\cite{RQ4-ExampleCleanup_a}$\rightarrow$\cite{RQ4-ExampleCleanup_b}), the renaming of \texttt{\seqsplit{libFilenameFilter$\rightarrow$LibFilenameFilter}} is associated with the message ``\textit{Lots of fixes using Checkstyle - Fixed some names to follow conventions...}''.

The Refactoring topic includes words such as `refactor', `revert', `updated', `changed', `removed, and `add'. These commits are associated with developers updating the code related to the behavior and design of the system. For example, the commit message of last rename in the chain: \texttt{\seqsplit{KenyaEmrConfigurator$\rightarrow$KenyaEmrModelConfigurator$\rightarrow$EmrModelConfigurator}} (\cite{RQ4-ExampleRefactor_a}$\rightarrow$\cite{RQ4-ExampleRefactor_b}) is ``\textit{Major refactor to start process of eventually moving content manager classes into separate module. For now they are moved to a different subpackage but remain in the KenyaEMR module until all dependencies on KenyaEMR are removed}''.

Finally, the Bug Fix/Testing topic is associated with the words `fix', `bug', `test', and `testcase'. In these instances, the renames are part of either a bug fix developers perform or are part of unit testing. However, we do notice that usually, the messages are not very descriptive. For example, the last message in the chain \texttt{\seqsplit{result$\rightarrow$dependencies$\rightarrow$calc}} (\cite{RQ4-ExampleFix_a}$\rightarrow$\cite{RQ4-ExampleFix_b}) is ``\textit{fixed bug with searching for transitive dependencies + added test for it}''.

The topics yielded from our analysis are at a high level. While they show the actions causing the rename, further insight into why the developer utilized a specific word for the rename or how the name is related to the action or code is challenging due to the nature of commit messages. 

\begin{tcolorbox}[top=0.5pt,bottom=0.5pt,left=1pt,right=1pt]
\textbf{Summary for RQ4.}
A topic modeling analysis on the rename chain commit messages shows the renames are related to Code Cleanup, Refactoring, and Bug Fix/Testing. However, these topics are at a high-level due to the nature of commit messages. 
\end{tcolorbox}

\section{Threats To Validity}
\label{Section:threats}
Even though the projects are limited to Java systems and might not necessarily generalize to systems written in other languages, these systems follow software engineering best practices and have been utilized in similar research on identifier names. Likewise, our methodology utilizes specific tools, such as RefactoringMiner and the part-of-speech tagger, which pose a risk because they could not be entirely accurate. These tools, however, are well-known and state-of-the-art in their respective domain and employed in similar work. Even though our construction of rename chains is limited to identifiers renamed within the same class, our results still yield a large number of chains. In RQ2, we utilize the commit author email to identify individual developers. While this can introduce threats to the study, manually verifying our dataset's large volume of emails is not feasible. Furthermore, as mentioned in RQ2, emails for identifying developers have been used in prior work.

\section{Discussion \& Conclusion}
\label{Section:discussion}
Interpreting identifier names form the backbone of any code comprehension task. However, with developers free to craft names using words of their choosing, they introduce the threat of having names that do not accurately reflect their behavior (i.e., names of poor quality), which hinders the maintenance of the system. To correct such poor-quality names, developers rename them, which can continue throughout the system's lifetime. In this study, our analysis of multiple renames applied to a single identifier (i.e., rename chain) shows that almost all projects exhibit this phenomenon, with an average chain size of two renames. Furthermore, we report on characteristics such as the interval between renames, developers responsible for chain construction, and grammatical changes. While our findings extend the knowledge in identifier naming, there are avenues for further research, including expanding on our RQ4 analysis to study the motivation and contextualization for the occurrence of rename chains. Below, we discuss how the findings from our RQs support the community through a series of takeaways.

\textbf{Takeaway 1 - \textit{Reliance on part-of-speech patterns when crafting and evaluating names.}}
From RQ3, we observe that part-of-speech tags are an efficient means of studying the semantic updates a name undergoes when renamed. This finding shows that academia and practitioners should not focus only on the words in a name but also consider the grammatical structure of the name when crafting and evaluating identifier names. Additionally, this also presents the research/vendor community with an opportunity to construct rename recommendation tools that incorporate the name's grammatical structure in addition to the existing features they utilize. 

\textbf{Takeaway 2 - \textit{Improvements to name recommendations and appraisal techniques.}}
In addition to incorporating the grammatical structure, identifier name recommendation and appraisal techniques should also consider the historical evolution of an identifier's name in the evaluation process. Current techniques usually consider the styling and features present in the version of the code base under analysis. By examining the historical evolution of the name, the likelihood of overreliance on outliers is greatly reduced.

\textbf{Takeaway 3 - \textit{Emphasis on the importance of using high-quality names.}}
Academia should instill in students the importance of having high-quality names in the source code. For example, our dataset shows the use of abbreviations and acronyms in forming identifier names. Such tokens are known to impede code comprehension  \cite{Hofmeister2017SANER, Schankin2018ICPC}. Specifically, the initial versions of the attribute rename chain: \texttt{\seqsplit{TEMP\_TUNNEL\_ID$\rightarrow$TUNNEL\_ID$\rightarrow$TUNNEL\_IDENTIFIER}} have a generic word, `TEMP', and an abbreviation, `ID', which are corrected in the final version of the name. Finally, in addition to using static analysis tools to detect poor programming practices, such as code and test smells \cite{Fernandes2016EASE,Aljedaani2021EASE}, there should also be a focus on using tools that evaluate the quality of names, such as linguistic anti-patterns \cite{Arnaoudova2016EMSE,Peruma2021ICSME}.

\textbf{Takeaway 4 - \textit{Challenges with the automated contextualization of rename chains.}}
Even though our attempts, in RQ 4, at contextualizing the occurrence of rename chains using the messages in the commit log yielded topics, these topics are at a high level. They are insufficient in helping us understand how the changed words in the identifier's name are related to the code or developer activity/task. This shows the need for more specialized natural language processing techniques and also the analysis of other software engineering artifacts.

\subsection{Future Work}
Our future work in this area includes a human subject study. In this proposed study, we will work with developers of varying experience and skills to validate our empirical findings and expand our knowledge on understanding the rationale for the presence of rename chains in projects. Further, we plan to discover additional heuristics we can incorporate into appraising and recommending high-quality identifier names.

\bibliographystyle{ieeetr}
\bibliography{main}

\end{document}